
\hfuzz=4pt
\magnification=1200

%
\catcode`\*=11  
\def\slashsym#1#2{\mathpalette{\sl*sh{#1}}{#2}}
\def\sl*sh#1#2#3{\ooalign{\setbox0=\hbox{$#2\not$}
                          $\hfil#2\mkern-24mu\mkern#1mu
                           \raise.15\ht0\box0\hfil$\cr
                          $#2#3$}}
\catcode`\*=12

\def\dslash{{\slashsym5\partial}}

\hsize=16truecm
\vsize=22truecm

\def\gsim{\mathrel{\raise.3ex\hbox{$>$\kern-.75em\lower1ex\hbox{$\sim$}}}}
\def\lesssim{\mathrel{\raise.3ex\hbox{$<$\kern-.75em\lower1ex\hbox{$\sim$}}}}

\baselineskip12pt
\rightline{UCLA/95/TEP/9}
\rightline{March 1995}
\bigskip
\bigskip
\bigskip
\bigskip
\bigskip

\centerline{\bf CHIRAL EFFECTIVE LAGRANGIAN DESCRIPTION}
\smallskip
\centerline{\bf OF BULK NUCLEAR MATTER}

\bigskip
\bigskip
\bigskip
\bigskip

\centerline{\bf Graciela Gelmini and Bruce Ritzi}
\centerline{Department of Physics, University of California, Los Angeles}
\centerline{Los Angeles, California 90024-1547}

\bigskip
\bigskip
\bigskip
\bigskip
\bigskip
\bigskip

\centerline{\bf ABSTRACT}
\bigskip
\bigskip
\bigskip

Here we point out that the four-nucleon terms incorporate into lowest
order non-linear chiral effective Lagrangians the same description of
bulk nuclear matter contained in the Walecka model, that is generally
considered satisfactory.  We find this point worth noticing because,
while the Walecka model is an
entirely phenomenological renormalizable model, non-linear chiral
Lagrangians have a deep basis in elementary particle physics, and, in
this sense, are more fundamental.

\vfill\eject

Chiral effective Lagrangians [1] are constructed on the basis of the
accidental global symmetries of QCD.
In particular, the approximate $SU_L(2) \times SU_R(2)$ global symmetry
is due to the smallness of the $u$ and $d$ quark masses with respect to
the confinement scale of QCD.
Its diagonal $SU_{\rm Axial}(2)$ symmetry is spontaneously broken at a
scale $\Lambda \lesssim 1 GeV$, and pions are the resulting
quasi-Nambu-Goldstone bosons, whose interactions, among themselves and
with fermions, are entirely determined by symmetry considerations at low
momenta $k < \Lambda$.
Thus, chiral effective Lagrangians have a deeper physical meaning than
other entirely phenomenological models of strong interactions.

\bigskip
In this letter we will use a non-linear $SU(2)\times SU(2)$ chiral
Lagrangian to describe the strong interactions of pions and nucleons.
We will apply this theory to bulk nuclear matter.
We want to point out that four-fermion terms incorporate into the chiral
Lagrangians the same description of bulk nuclear matter contained in the
Walecka model [2].
This description is the main feature of the Walecka model, that is
considered a satisfactory phenomenological model of nuclear matter.
Surface effects would, however, be different in both models, as would be
higher order corrections beyond the mean field approximation.
A previous attempt [3] to describe nuclei using non-linear $SU(2) \times SU(2)$
chiral Lagrangian incorporated a pion condensate as an essential feature.
Here, we do not assume pion condensation.

\bigskip
To describe nuclear matter, we use here the most general Lagrangian
describing the interactions of pions $\vec \pi = (\pi_1, \pi_2,\pi_3)$
and nucleons $N^T = (p,n)$, in which the spontaneously broken $SU(2)
\times SU(2)$ chiral symmetry is realized non-linearly.
This Lagrangian contains non-renormalizable terms of arbitrary high
dimensions, as is expected in an effective theory.
However, at small momenta of pions and nucleons, $k < \Lambda$, only a few
terms are relevant,
since higher dimensional terms are weighted down by powers of
$\Lambda^{-1}$ [4].
The lowest order terms in powers of $k / \Lambda$ and
 $m_\pi / \Lambda$ ($m_\pi$ is the pion mass ) are [5]
$$
\eqalign{
{\cal L} & = {1\over 2} D^{-1} \partial_\mu \vec\pi ~\partial^\mu \vec\pi -
{1\over 2} D^{-1} m_\pi^2 \vec\pi^2
+ \bar N \left [ \dslash - m_N + i D^{-1} f_\pi^{-1}
\gamma_5 g_A~ \vec t\cdot \dslash\vec\pi + \right.\cr
&\cr
& \left. + {i\over 2} D^{-1} f_\pi^2 ~\vec t\cdot(\vec\pi \times
\dslash\vec\pi)\right ] N + {\cal L}_{4-N}~~.\cr}
\eqno(1)
$$
Here $\vec t$ are the isospin generators, $\vec t = \vec\sigma/2$
($\vec\sigma$ are the pauli matrices), $g_A
\simeq 1.25$ and $f_\pi = 93 MeV$ are the axial coupling and pion
 decay constants respectively, $D\equiv 1 +
\vec\pi^2/4f_\pi^2$, and the four-fermion Lagrangian ${\cal L}_{4-N}$ is
$$
{\cal L}_{4-N} ={ C_\alpha^2\over 2f_\pi^2}(\bar N \Gamma_\alpha N) (\bar
N \Gamma^\alpha N)~~.
\eqno(2)
$$
Where $\Gamma_\alpha$ stands for $1$, $\gamma_5$, $\gamma_\mu$,
$\gamma_\mu\gamma_5$ and $\sigma_{\mu\nu}$, in principle multiplied or not by
$\vec t$. However,
the terms which include a $\vec t$ can be rewritten, using a Fierz
reordering, as linear combinations of terms without a $\vec t$. Moreover,
most of these terms vanish in bulk nuclear matter.

In the static, uniform system of bulk matter the three vector momentum
and spin dependent interactions average to zero due to rotational symmetry.
This is easily seen by picturing each single baryon acted  upon by
the two body forces of  all the remaining baryons in the media, that
 occupy the entire set of states in a fermi sphere, therefore
including all directions of momentum for a given energy, as well as all
spin orientations. Thus, the sum of these forces on each
single baryon vanishes.
The spin and three vector momentum dependent interactions are those
for which $\Gamma_\alpha = \gamma_5$, $\gamma_i$,
 $\gamma_\mu\gamma_5$, and $\sigma_{\mu\nu}.$
Therefore, only $\Gamma_\alpha = 1$, $ \gamma_0$ remain in Eq. (2), namely
$\bar NN$ and $N^\dagger N$. In the non-relativistic
limit the difference of these two operators is of order $k^2/m_N^2$
($m_N$ is the nucleon mass).
Nucleons in nuclear matter are non-relativistic, however $\bar NN$ and
$N^\dagger N$ differ by approximately 10$\%$, therefore we keep both terms,
$$
{\cal L}_{4-N} = {C_S^2\over 2f_\pi^2} (\bar NN)(\bar NN) -
{C_V^2\over 2f_\pi^2}
(N^\dagger N)(N^\dagger N)~~.
\eqno(3)
$$
The fact that these two terms, and only these two, remain from Eq. (2)
in bulk nuclear matter, is what makes the equivalence with the Walecka
model possible. It is also worth noting that higher order terms in
$\bar NN$ or $N^\dagger N$ can be constructed by adding $\bar
NN/f_\pi^2\Lambda$
or  $(N^\dagger N/f_\pi^2\Lambda)^2$ [4]
to terms already present in the chiral lagrangian, but since
$\bar NN \simeq N^\dagger N = \rho_B$ (see Eq.8 below),
the baryonic density, that is here the
nuclear density $\rho_B = \rho_N$, and $f_\pi^2\Lambda \simeq 7\rho_N$,
 these terms are less important and will not be considered in this letter.

\bigskip
There is a point that needs clarification  when using nucleons in chiral
Lagrangians.
We said we are using an expansion in small factors weighted down by
 $\Lambda^{-1}$. However, the time derivative of $N$ gives a factor
 of order $m_N$, that is not small compared to $\Lambda$.
Thus, terms containing only powers of $\partial_0N$ are large
and should be summed up.
However, we can eliminate each nucleon time derivative  in
the interaction Lagrangian in favor of only small factors by using
Dirac's equation [5].
After this substitution by space derivatives and other small factors,
the remaining terms have the same form as terms already present in the
interaction Lagrangian (i.e. those having no time derivatives
of the nucleon fields), and simply renormalize them (this procedure
is equivalent to  a redefinition of the nucleon fields [6]).
Thus the only time derivative that remains is in the non-interacting
part of the Lagrangian, given in Eq. (1).

\bigskip
Using the Lagrangian in Eqs. (1) and (3), we can now solve for the energy
density $\epsilon$ and pressure $P$ of nuclear matter, and determine the
only relevant parameters, $C_V$ and $C_S$ by reproducing the binding
energy $(\epsilon/\rho_B-m_N)$ and the density $\rho_B$ of nuclear
matter [2].
We treat nucleons as a Fermi gas, in the mean field generated by all
nucleons due to the effective interactions in Eq. (3). There is no net
pion mean field in nuclei, and the effects of pion exchanges
average essentially to zero in the description of the bulk properties of
nuclei, due to the spin dependence of the pion-nucleon coupling.
Therefore, all the terms containing pions in Eq. (1) do not contribute. From
Dirac's equation,
$$
\left [i \gamma^\mu\partial_\mu - m_N - {C_V^2\over f_\pi^2}
N^\dagger N\gamma_0 + {C_S^2\over f_\pi^2} \bar NN\right ] N = 0~~.
\eqno(4)
$$
we can recognize an effective mass by summing the scalar terms in the
bracket,
$$
m_* = m_N - {C_S^2\over f_\pi^2} \bar NN~~,
\eqno(5)
$$
while the $\gamma_0$ dependent $C_V^2$ term shifts the energy, so that
$E = \sqrt{k^2+m_*^2} + (C_V^2/f_\pi ^2)N^\dagger N$ (this can be seen by
writing the dispersion relation).
Rewriting the fermion Lagrangian in terms of $m_*$ and computing the
energy-momentum tensor $T^{\mu\nu}$, we get from their time and space
components
$$
\epsilon = T^{00} = \bar N i \gamma^i \partial_i N + m_*\bar NN+
{C_V^2\over f_\pi^2} N^\dagger N~N^\dagger N + {C_S^2\over f_\pi^2}\bar
 N~N\bar NN
\eqno(6)
$$
and
$$
P = {T^{ii}\over 3} = {1\over 3} \bar N i\gamma^i\partial_i N +
{C_V^2\over f_\pi^2} N^\dagger N~N^\dagger N - {C_S^2\over f_\pi^2} \bar
 N~N\bar NN~~.
\eqno(7)
$$

By filling up the Fermi sea up to the Fermi momentum $k_F$ in the ground
state $|N_0\rangle$ we get
$$
\rho_B = \langle N_0|N^\dagger N|N_0\rangle
= \gamma \int {d^3k\over (2\pi)^3} = {2k_F^3\over 3\pi^2}
\eqno(8)
$$
and
$$
\rho_S = \langle N_0|\bar NN|N_0\rangle =
\gamma\int {d^3k\over (2\pi)^3}
{m_*\over{\sqrt{k^2+m_*^2}}}~~,
\eqno(9)
$$
where $\gamma=4$ counts protons and neutrons and their spins.
In bulk nuclear matter, we can thus replace $N^\dagger N$ and $\bar NN$
in the Eqs. (6) and (7) by their values in the ground state of the
Fermi gas, $\rho_B$ and $\rho_S$ respectively.
Moreover for free nucleons of momentum $k$, $\bar N\gamma^iN = k_i/E$,
thus
$\bar N\gamma^i\partial_i N=k^2/E$ and in the ground state Eqs.(6) and
(7) become,
$$
\epsilon = \int {d^3k\over (2\pi)^3}~~ {\sqrt{k^2 + m_*^2} }
+ {C_V^2\over f_\pi^2} \rho_B^2 +
{C_S^2\over f_\pi^2} \rho_S^2
\eqno(10)
$$
and
$$
P = \int {d^3k\over (2\pi)^3}~~
{k^3\over 3{\sqrt{k^2 + m_*^2}} }
+ {C_V^2\over f_\pi^2} \rho_B^2 -
{C_S^2\over f_\pi^2} \rho_S^2~~.
\eqno(11)
$$
Eqs. (10) and (11) exactly reproduce the results of the Walecka model
for bulk nuclear matter.

\bigskip
The Walecka model [2] introduces a neutral scalar meson field $\phi$ of mass
$m_\phi$ and a neutral vector meson field $V_\mu$ of mass $m_V$ with
couplings
$$
\bar N(g_\phi \phi - g_V \gamma^\mu V_\mu)N
\eqno(12)
$$
with the nucleons.
In bulk nuclear matter where $\rho_B$ and $\rho_S$ are constant, $\phi$
and $V_o$ become constants and $V_i = 0$.
Nucleons are treated as a Fermi gas in the mean fields  $\phi$ and $V_o$,
whose values
$$
\phi = {g_\phi\over m_\phi^2} \bar NN \quad,\quad V_o =
{g_v\over m_v^2} N^\dagger N~~,
\eqno (13)
$$
are obtained by solving the classical equations of motion.
By comparing $m_*$, $\epsilon$ and $P$ obtained in this way with the
Walecka model, with the chiral Lagrangian results in
Eqs. (5), (10) and (11), we see that they coincide if the parameters of both
Lagrangians are related by
$$
{C_V^2\over f_\pi^2} = {g_V^2\over m_V^2} \quad, \quad
{C_S^2\over f_\pi^2} = {g_\phi^2\over m_\phi^2}~~.
\eqno(14)
$$

Even if the Walecka Lagrangian seems to have more free parameters than
the four fermion interaction in Eq. (3), we see that the number of
independent parameters are the same.
By comparison with the results of Chin and Walecka [2] we obtain that
$$
C_V = 1.34\quad,\quad C_S = 1.57\quad,
\eqno(15)
$$
fit the binding energy and density of bulk nuclear matter.

\bigskip
The equivalence of the results of both models in bulk nuclear matter
becomes obvious, if one realized in this medium,
where $\rho_B$ and $\rho_S$ are
constant, the Lagrangian of the Walecka model reduces to
$$
{\cal L}_W = \bar N (i\dslash-m_N) N +
{g_\phi^2\over 2m_\phi} \bar NN~\bar NN - {g_v^2\over 2m_\phi^2}
N^\dagger N~N^\dagger N~~,
\eqno(16)
$$
when the Eqs. (13) are used.
Using the relations among parameters in Eq. (14), we see that this is
exactly the form of the chiral Lagrangian we applied to the bulk media.

\bigskip
It is interesting to compare the values of $C_V$ and $C_S$ obtained
here to account for  bulk nuclear properties to those derived using
the chiral Lagrangian, Eqs.(1) and (2), to describe low energy
nucleon-nucleon scattering [5].
At the low energy considered in these scattering $E< 5 MeV$, the difference
between $\bar NN$ and $N^\dagger N$ is negligible, thus only one
parameter, $C_S^2 - C_V^2$
 (that is $f_\pi^2C_s$ in Weinberg's notation [5])
enter in the Lagrangian and is determined [5,7,8] to be
$$
C_S^2 - C_V^2 = {f_\pi^2\over (88MeV)^2} = 1.12\quad.
\eqno(17)
$$
Here we obtain $C_S^2 - C_V^2 = 0.67$ instead, using Eq. (15). It
is expected here to obtain a lower value, because repulsive forces
are known to be more important in nuclear matter than in the low
energy scattering processes considered above [8].

\bigskip
We have shown here that the four-nucleon terms incorporate into chiral
Lagrangians the same description of bulk nuclear matter of the Walecka
 model, in the mean field approximation. However,
 surface effects and corrections beyond the mean field approximation
should be different and worth exploring.
\vfill\eject

\noindent
{\bf References}
\bigskip
\bigskip

\item{[1]} S. Weinberg, Phys. Rev. Lett. {\bf 18} (1967) 188.

\item{ } S. Weinberg, Phys. Rev. {\bf 166} (1968) 1568.

\item{ } S. Weinberg, Physica A {\bf 96} (1979) 372.
\bigskip

\item{[2]} S. A. Chin and J. D. Walecka, Phys. Lett. {\bf 52B}
(1974) 14.
\bigskip

\item{[3]} B. Lynn, Nucl. Phys. {\bf B402} (1993) 281.
\bigskip

\item{[4]} Manohar and Georgi, Nucl. Phys. {\bf B234} (1984) 189.
\bigskip

\item{[5]} S. Weinberg, Phys. Lett. {\bf B251} (1990) 288.
\item{} S. Weinberg, Nucl. Phys. {\bf B363} (1991) 3.

\bigskip
\item{[6]} E. Jenkins and A.V. Manohar, Phys. Lett. B255 (1991) 558.

\bigskip
\item{[7]} D. Montano, H.D. Politzer, and M.B. Wise, Nucl. Phys.
{\bf B375} (1992) 507.

\bigskip
\item{[8]} A. Bohr and B.R. Mottleson, Nuclear Structure, Vol. 1
(Benjamin, New York, 1969)

\end

\bye